\documentclass[epj]{svjour}
\usepackage{latexsym}
\usepackage{graphics}
\usepackage{graphicx}
\usepackage{amssymb}   
\usepackage{url}
\usepackage{amsmath}
\usepackage{aas_macros}
\begin{document}
\title{Neutron capture measurements for s-process nucleosynthesis}
\subtitle{A review about CERN n\_TOF developments and contributions}
%
\offprints{}          
\institute{Insert the first address here, the second here}
\date{Received: date / Revised version: date}
%
\authorrunning{C. Domingo-Pardo et al.}

\author{\footnotesize
C.~Domingo-Pardo$^{1}$, %
O.~Aberle$^{2}$, %
V.~Alcayne$^{3}$, %
G.~Alpar$^{4}$, %
M.~Al~Halabi$^{5}$, %
S.~Amaducci$^{6}$, %
V.~Babiano$^{7}$, %
M.~Bacak$^{2,8}$, %
J.~Balibrea-Correa$^{1}$, %
J.~Bartolom\'{e}$^{9}$
% ...
The n\_TOF Collaboration (\url{www.cern.ch/ntof})
}

\institute{\footnotesize
$^{1}$Instituto de F\'{\i}sica Corpuscular, CSIC - Universidad de Valencia, Spain
$^{2}$European Organization for Nuclear Research (CERN), Switzerland
$^{3}$Centro de Investigaciones Energ\'{e}ticas Medioambientales y Tecnol\'{o}gicas (CIEMAT), Spain
$^{4}$University of Dallas, USA
$^{5}$Goethe University Frankfurt, Germany
$^{6}$INFN Laboratori Nazionali del Sud, Catania, Italy
$^{7}$Universitat Polit\`{e}cnica de Catalunya, Spain
$^{8}$University of Manchester, United Kingdom
$^{9}$Universidad de Sevilla, Spain
$^{10}$CEA Irfu, Universit\'{e} Paris-Saclay, F-91191 Gif-sur-Yvette, France
$^{11}$Helmholtz-Zentrum Dresden-Rossendorf, Germany
$^{12}$Horia Hulubei National Institute of Physics and Nuclear Engineering, Romania
$^{13}$Department of Physics, Faculty of Science, University of Zagreb, Zagreb, Croatia
$^{14}$University of Santiago de Compostela, Spain
$^{15}$Agenzia nazionale per le nuove tecnologie, l'energia e lo sviluppo economico sostenibile (ENEA), Italy
$^{16}$Istituto Nazionale di Fisica Nucleare, Sezione di Bologna, Italy
$^{17}$Istituto Nazionale di Fisica Nucleare, Sezione di Trieste, Italy
$^{18}$Department of Physics, University of Trieste, Italy
$^{19}$INFN Laboratori Nazionali di Frascati, Italy
$^{20}$Istituto Nazionale di Fisica Nucleare, Sezione di Torino, Italy
$^{21}$Department of Physics, University of Torino, Italy
$^{22}$Istituto Nazionale di Fisica Nucleare, Sezione di Bari, Italy
$^{23}$Istituto Nazionale di Fisica Nucleare, Sezione di Perugia, Italy
$^{24}$Istituto Nazionale di Astrofisica - Osservatorio Astronomico d'Abruzzo, Italy
$^{25}$National Technical University of Athens, Greece
$^{26}$Physikalisch-Technische Bundesanstalt (PTB), Bundesallee 100, 38116 Braunschweig, Germany
$^{27}$University of Ioannina, Greece
$^{28}$University of York, United Kingdom
$^{29}$Affiliated with an institute covered by a cooperation agreement with CERN
$^{30}$University of Granada, Spain
$^{31}$University of Lodz, Poland
$^{32}$TU Wien, Atominstitut, Stadionallee 2, 1020 Wien, Austria
$^{33}$Istituto Nazionale di Fisica Nucleare, Sezione di Roma1, Roma, Italy
$^{34}$European Commission, Joint Research Centre (JRC), Geel, Belgium
$^{35}$Charles University, Prague, Czech Republic
$^{36}$School of Physics and Astronomy, University of Edinburgh, United Kingdom
$^{37}$Dipartimento di Fisica e Astronomia, Universit\`{a} di Bologna, Italy
$^{38}$INFN Laboratori Nazionali di Legnaro, Italy
$^{39}$Dipartimento Interateneo di Fisica, Universit\`{a} degli Studi di Bari, Italy
$^{40}$Paul Scherrer Institut (PSI), Villigen, Switzerland
$^{41}$Consiglio Nazionale delle Ricerche, Bari, Italy
$^{42}$Istituto Nazionale di Fisica Nucleare, Sezione di Catania, Italy
$^{43}$Department of Physics and Astronomy, University of Catania, Italy
$^{44}$Instituto Superior T\'{e}cnico, Lisbon, Portugal
$^{45}$Japan Atomic Energy Agency (JAEA), Tokai-Mura, Japan
$^{46}$Department of Physics and Astronomy, Uppsala University, Box 516, 75120 Uppsala, Sweden
}


%
%

\date{Received: \today / Revised version: \today}

\abstract{
This article presents a review about the main CERN n\_TOF contributions to the field of neutron-capture experiments of interest for $s$-process nucleosynthesis studies over the last 25 years, with special focus on the measurement of radioactive isotopes. A few recent capture experiments on stable isotopes of astrophysical interest are also discussed. Results on $s$-process branching nuclei are appropriate to illustrate how advances in detection systems and upgrades in the facility have enabled increasingly challenging experiments and, as a consequence, have led to a better understanding and modeling of the $s$-process mechanism of nucleosynthesis. 
New endeavors combining radioactive-ion beams from ISOLDE for the production of radioisotopically pure samples for activation experiments at the new NEAR facility at n\_TOF are briefly discussed. 
On the basis of these new exciting results, also current limitations of state-of-the-art TOF and activation techniques will be depicted, thereby showing the pressing need for further upgrades and enhancements on both facilities and detection systems. A brief account of the potential technique based on inverse kinematics for direct neutron-capture measurements is also presented. 
\PACS{
      {PACS-key}{discribing text of that key}   \and
      {PACS-key}{discribing text of that key}
     } 
} 
\maketitle
\section{Introduction}
\label{intro}
Neutron-capture reactions play a fundamental role in the cosmic origin of elements heavier than iron, both during hydrostatic stages of stellar evolution ($s$-process) and in cataclysmic stellar environments ($r$-process). 
These two nucleosynthesis mechanisms were first presented in detail in the works of B2FH \cite{BBFH} and Cameron \cite{Cameron57}. As a consequence, a large experimental effort over the last 70 years has resulted in a wealth of neutron-capture nuclear data. The first experiments were initially aimed at validating the $s$-process hypothesis \cite{Macklin62,Macklin63c}, and shortly afterwards to aid in the development and refinement of stellar models while  constraining the physical conditions along different evolutionary stages of stars \cite{Macklin67,Gibbons67}. See also the review of K\"appeler et al. \cite{Kaeppeler11} and references therein. 

Two different methodologies for neutron-capture cross section measurements have been extensively applied so far in many laboratories worldwide, neutron time-of-flight (TOF) and neutron activation, thereby covering about 350 (predominantly stable) nuclei \cite{Dillmann23}. However, there are still acute needs for new neutron-capture cross-section measurements and for a large fraction of the measured isotopes improvements are required both in terms of accuracy and energy-range completeness. 

From the astrophysical standpoint there are three “families” of nuclides with a particular value for $s$-process studies. The group of s-only nuclei serves as a benchmark for $s$-process and galactic-chemical evolution (GCE) models, which should reproduce 100\% of the $s$-only isotopic abundances \cite{Travaglio04,Bisterzo15,Prantzos20}. The group of $s$-process bottlenecks gives rise to the three characteristic $s$-process abundance peaks at the neutron-shell closures N = 50, 82 and 126. Because of their large abundances, elements in the bottleneck peaks show-up prominently in spectroscopic observations of stellar atmospheres and thus, they represent a sensible probe for stellar models. Finally, $s$-process branching nuclei are especially relevant for constraining the physical conditions of the stellar environment \cite{Kaeppeler11}. Interestingly, for none of these groups the quality of the data complies with the $\pm$5\% uncertainty level required by stellar models. 
The neutron-capture cross sections of most s-only nuclei are relatively well determined \cite{Dillmann23}, which does not necessarily imply that their s-process abundances are correspondingly accurate. This is a consequence of the strong interplay between many s-process branchings and s-only isotopes \cite{Bisterzo14,Bisterzo15}. The astrophysical impact of the uncertainties on the s-process branchings is therefore twofold, because the final s-only abundances are important both for benchmarking the performance of stellar models \cite{Arlandini99} and for deriving the isotopic $r$-process abundance distributions in the solar system \cite{Goriely99,Arnould07,Prantzos20}. 

Past and ongoing efforts at CERN n\_TOF to improve this situation are described in this article. A large number of articles describe already the n\_TOF facility and the related measuring technics in great detail \cite{Rubbia98,Gunsing11,Guerrero13,Colonna18b,Domingo23a} and thus, only some main facility features relevant for the discussions in the present article will be summarized in Sec.\ref{sec:ntof}. Because of their relevance for $s$-process nucleosynthesis and the demanding features of the experiments, measurements on unstable $s$-process branching nuclei are well suited to illustrate the experimental progress achieved over the last 25 years. Thus, Sec.\ref{sec:branchings} presents in chronological order the main results on $s$-process branching nuclei measured at n\_TOF, along with the facility and detector advances that were relevant for such studies.
Measurements on stable isotopes involved in the $s$-process path are also of paramount importance for properly interpreting observed elemental abundances in stars and isotopic analysis of meteorites, and thus for a better understanding of the $s$-process mechanism. Some recent examples will be discussed in Sec.\ref{sec:stable}.
Finally, Sec.\ref{sec:prospects} summarizes some of the main limitations of state-of-the-art TOF experiments, and presents ongoing efforts to improve present instruments and to complement them with new measuring stations and techniques.

\section{The n\_TOF facility}\label{sec:ntof}
\begin{figure*}[!htbp]
\centering
\includegraphics[width=2\columnwidth]
{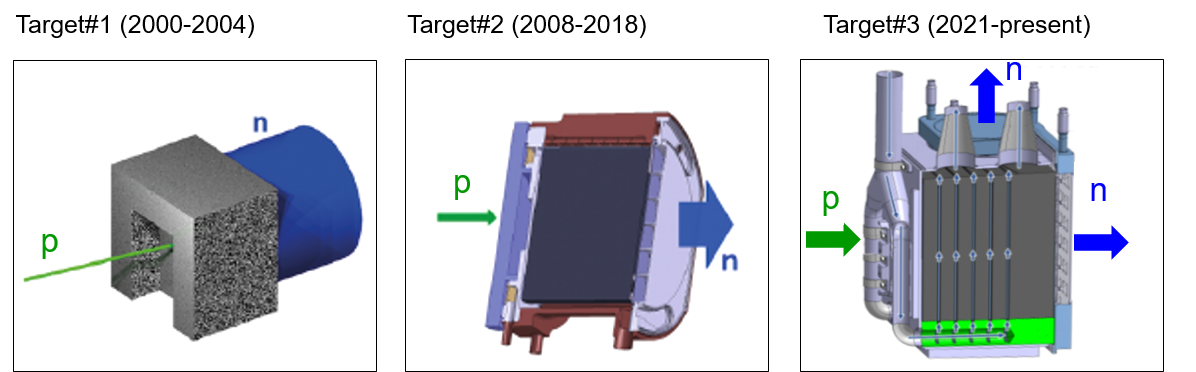}
\caption{From left to right, the three generations of spallation sources at n\_TOF. The latest version \cite{Esposito21} delivers a superior neutron-beam quality for capture experiments at both measuring stations EAR1 (185 m) and EAR2 (20 m) \cite{Lerendegui23a}.}
\label{fig:target}       
\end{figure*}

n\_TOF utilizes a 6 ns wide, 20 GeV pulsed proton beam from CERN’s Proton Synchrotron (PS), typically delivering nowadays 8$\times$10$^{12}$ protons per pulse to a lead spallation target (Fig.\ref{fig:target}), where approximately 300 neutrons are produced per incident proton. The neutron energy is determined with a high precision by means of the time-of-flight technique \cite{Lorusso04}. The low duty cycle and the high instantaneous flux are two of the most remarkable facility features. Neutron bunches are spaced at intervals of at least 1.2 seconds, corresponding to the operational cycle of the CERN Proton Synchrotron (PS). This duty cycle enables measurements over an extended TOF span, facilitating the detection of low-energy neutrons without overlap from subsequent neutron cycles. As a result, neutron energies as low as approximately 10 meV can be measured, with the high-energy portion of the spectrum remaining unaffected by slow neutrons from previous cycles. On the other hand, the exceptionally high instantaneous neutron flux is particularly advantageous when working with radioactive samples, as it ensures a highly favorable ratio of neutron-induced reaction signals to background signals caused by radioactive decay events.

 At the core of the facility, the optimization of the spallation target has been one of the key aspects for the successful realization of neutron-capture experiments (see Fig. \ref{fig:target}). A layer of water moderates the initially fast neutrons into a white spectrum covering a broad energy range from meV to GeV, thus fully including the energy-range of astrophysical interest for $s$-process studies ($\sim$1 eV - 100 keV). During phase-I (2001-2004) a water layer served as coolant and moderator for a 80$\times$80$\times$60 cm$^3$ lead target (Fig.\ref{fig:target}). This design severely limited the lifetime of the target owing to corrosion effects and, more relevant for capture experiments, contaminant neutron-capture reactions in the hydrogen of the water induced a large in-beam $\gamma$-ray background. The latter was a minor issue for the measurement of isotopes with large capture cross sections, like unstable $^{151}$Sm \cite{Marrone04}, but it made quite difficult the measurement of neutron-magic isotopes with small capture cross sections, like $^{208}$Pb or $^{209}$Bi \cite{Domingo06a}. A new cylindrical (40 cm length, 60 cm diameter) target design (see Fig.\ref{fig:target}) incorporated several improvements for phase-II (2009-2012), among them the use of a 1 cm water cooling circuit and a separate $^{10}$B-loaded 4-cm thick demineralized-water moderator, which efficiently suppressed the in-beam (E$_\gamma$=2.2 MeV) $\gamma$-ray background. However, during the last years of operation an increase in the activity of the cooling circuit was detected, which was ascribed to corrosion in the lead-water interface. In 2014 a second experimental area EAR2 was constructed at only 20 m above the already existing spallation target \cite{Chiaveri12,Barros15}. The new measuring extension allowed one to significantly extend the number of experiments during phase-III (2014-2018) (see Fig.1 in Ref.\cite{Domingo23a}). Initially, the large instantaneous flux of EAR2 could not be efficiently exploited for $s$-process capture studies due to the fact that the target geometry and moderator characteristics were not optimal for the new EAR2 station \cite{Lerendegui16,Lomeo15}. However, all these limitations were remarkably improved in phase-IV (2021-) with the third-generation neutron target \cite{Esposito21} (see Fig.\ref{fig:target}). With the new target a slightly higher neutron flux was achieved and, more importantly, the resolution function (RF) at EAR2 could be significantly enhanced (see Fig.2 in \cite{Lerendegui23a}). A large instantaneous neutron flux and a narrow resolution-function become of special importance for astrophysics experiments \cite{Koehler00} and thus, the new target features enabled at EAR2 some of the most challenging and fascinating neutron-capture TOF experiments performed so far, as it is discussed below.

\section{Progress on $s$-process branching isotopes}\label{sec:branchings}
Unstable nuclei with relatively long half-lives ($\sim$yr) are particularly interesting for $s$-process studies because they produce a split in the $s$-process path. The strenth of the branching determines the local isotopic pattern around the unstable nucleus and the resulting isotopic abundances become sensitive to the physical conditions of the stellar environment \cite{Kaeppeler11}. Therefore, accurate neutron-capture cross section measurements on these nuclides in combination with isotopic analysis of meteorites and state-of-the-art stellar models may provide correspondingly valuable constraints on the thermal- and neutron-density conditions of the stellar environment. 
From the experimental viewpoint, producing (radioactive) samples with sufficient number of atoms ($\geq 10^{18}$ atoms), and with high enough enrichment becomes a true challenge. For most of the experiments discussed in this section a sample-production strategy based on the combined effort from Institut Laue-Langevin (ILL)-Grenoble (France) and Paul Scherrer Institute (PSI)-Villigen (Switzwerland) was pursued. At ILL isotope-production via thermal-neutron activation was carried out, while the radiochemistry laboratory of PSI produced the samples for irradiation at ILL and in some cases performed a posterior radiochemical separation and purification. Still, many of the measured radioactive samples posed important challenges for the capture experiment, which were mainly related to the sample radioactivity, to the final number of atoms available for the isotope of interest and to the sample purity or enrichment. In order to cope with such demanding cases, a big effort has been made over the last 25 years at CERN n\_TOF for optimizing the quality of the neutron beam in terms of resolution function, high instantaneous neutron flux and low gamma- and neutron-induced backgrounds. In addition, both continuous adaptations and disruptive approaches have been pursued with the detection systems, leading to a progressive increase in detection sensitivity for the radiative neutron-capture channel of astrophysical interest. These advances are described in more detail below.
\begin{figure*}[!htbp]
\centering
\includegraphics[width=2\columnwidth]
{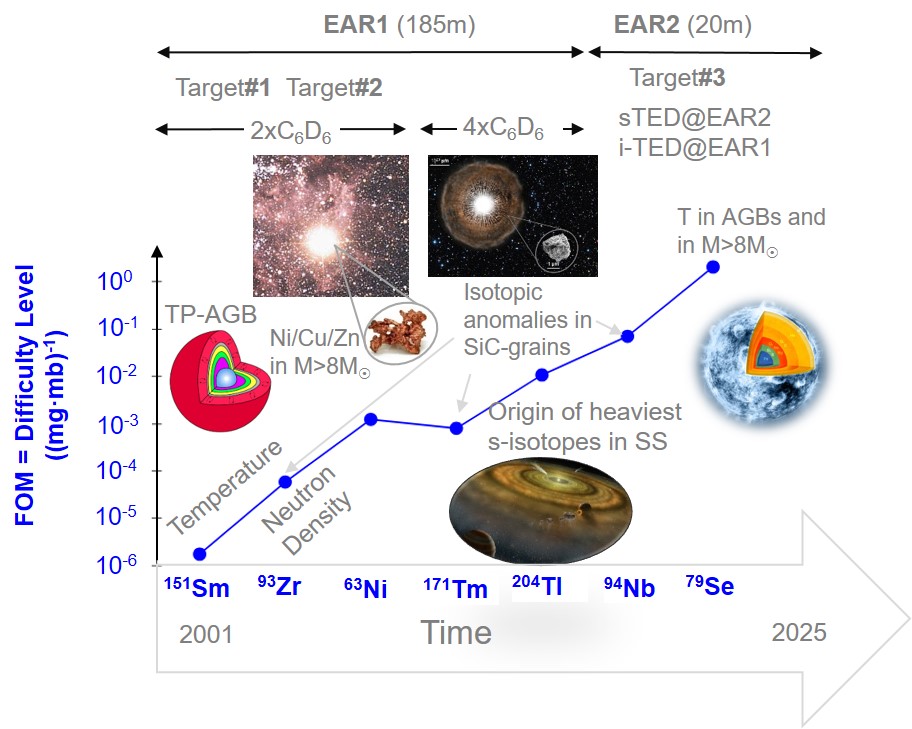}
\caption{Advances in facility upgrades and detection systems at CERN n\_TOF have led to increasingly difficult experiments on $s$-process branching nuclei along the last two decades. See text for details on the different $s$-process branching cross-section measurements.  For Target\#1-3 see Fig.\ref{fig:target} and for the detection system refer to Fig.\ref{fig:ear1} and Fig.\ref{fig:ear2}. Credits for figure insets: TP-AGB drawing adapted from \cite{Lugaro23};  HST image of the TP-AGB U. Camelopardalis with a SiC grain (NASA/Nan Liu/Andrew Davis), massive-star illustration (NASA/CXC/M. Weiss).}
\label{fig:sensitivity}       
\end{figure*}
Such progress can be well illustrated via the measured $s$-process branching isotopes and the corresponding facility and detector upgrades. To do so, a figure of merit (FOM) is arbitrarily defined here as 
\begin{equation}\label{eq:fom}
FOM = \frac{1}{m_i \cdot \sigma_i \cdot f_e},
\end{equation}

where $m_i$ is the mass of the isotope of interest in the sample (in mg), $\sigma_i$ is the Maxwellian capture cross section at 30 keV (in mb), and $f_e$ is the enrichment factor for the isotope of interest in the sample. This FOM will be also referred to as "difficulty level"  along this article, and it is shown in Fig. \ref{fig:sensitivity} as a function of the $s$-process branching isotope measured (published) along the different years. In short, the smaller are the sample mass, the cross section and the enrichment, the more difficult it becomes to perform the measurement. The FOM does not include the half-life of the isotope, nor the difficulty ascribed to the background induced by the sample-decay itself, which can be very different depending on whether there is emission of high-energy $\gamma$-rays or not. However, as we will see, for the cases and the range of half-lives discussed here (2 yr -- 4 Myr) there is no correlation between difficulty and half-life, because the other ingredients became a much more significant constraint for the measurement. Remarkably, as it will be seen later, at least for the cases described here the main limitations in terms of sample-radioactivity were not due to the decay of the isotope of interest itself, but rather to different levels of $^{60}$Co contamination in the samples (see discussion below for $^{204}$Tl, $^{94}$Nb and $^{79}$Se).

A more relevant ingredient, which is not included in the FOM, is the effective neutron-energy range actually covered in the measurement or the neutron-energy range, which is usable in the analysis of the capture data. This aspect is difficult to include in such a FOM and, as we will see, it is worth recognizing that an effort should be made to extend the energy range for most of the measured $s$-process branching points discussed here, once alternative approaches or improved measuring conditions become available in the future.

\subsection*{$^{151}$Sm($n,\gamma$): temperature in He-shell flashes of low-mass red giant stars}
\begin{figure*}[!htbp]
\centering
\includegraphics[width=2\columnwidth]
{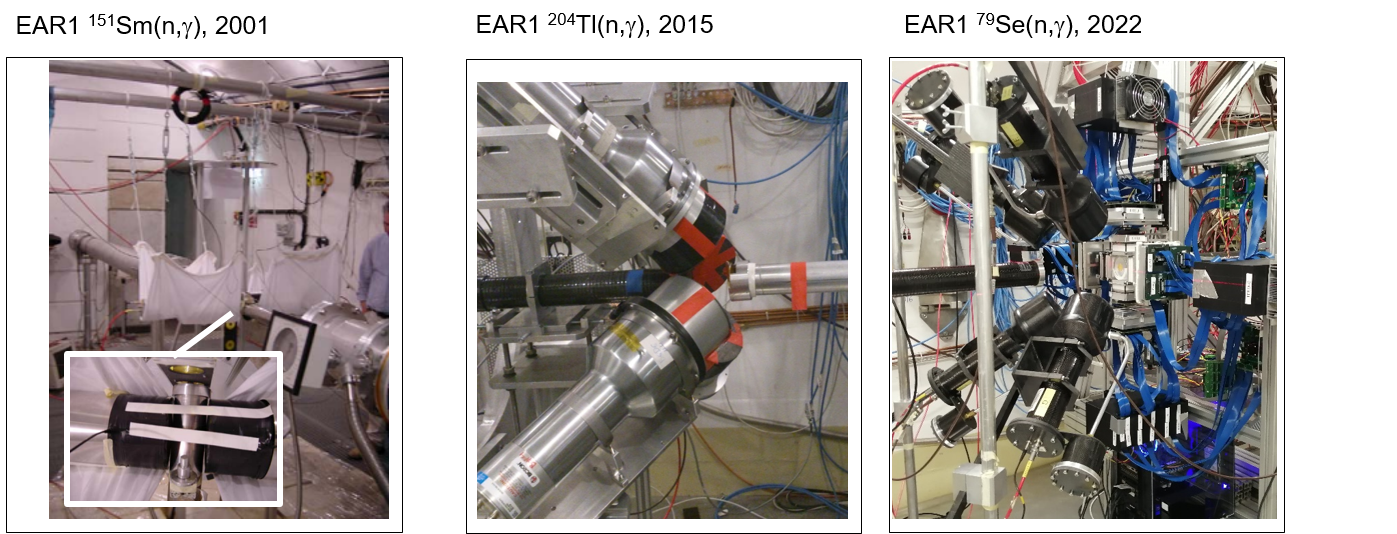}
\caption{Evolution of the capture set-up at EAR1, from left to right, two first-generation C-fiber based C$_6$D$_6$ detectors for the measurement of $^{151}$Sm($n,\gamma$) \cite{Marrone04}, four Bicron C$_6$D$_6$ detectors with lead shields for the $^{204}$Tl($n,\gamma$) experiment \cite{Casanovas24b} and four latest-generation C-fiber C$_6$D$_6$ detectors \cite{Mastinu21} plus the i-TED Compton array \cite{Domingo16} for the $^{79}$Se($n,\gamma$) cross-section measurement \cite{Lerendegui23b}.}
\label{fig:ear1}       
\end{figure*}

$^{151}$Sm (t$_{1/2}$ = 94.6 y) was the very first $s$-process branching measured at the beginning of the n\_TOF experiment in 2001 and published in 2004 \cite{Marrone04}. A highly enriched ($\sim$90\%) sample of $^{151}$Sm with a mass of 206 mg (8$\times$10$^{20}$ atoms) was measured at the EAR1 station using a set of two C$_6$D$_6$ detectors placed at 90$^{\circ}$ with respect to the beam axis at the sample position (see Fig.\ref{fig:ear1}). These detectors where designed with the aim of enhancing detection efficiency and reducing neutron-induced backgrounds \cite{Plag03} (see also discussion in Sec.\ref{sec:stable}). Both in-beam $\gamma$-rays background and the high sample activity (156 GBq) played a minor role over the entire energy range of interest for this measurement (see Fig.2 in \cite{Marrone04}) due to the large capture cross section ($\sim$3 b at 30 keV) and the fact that $^{151}$Sm decays via pure $\beta$ emission without high-energy $\gamma$-rays. 
Despite being the very first TOF measurement of the neutron-capture cross section of $^{151}$Sm it was quite straight-forward and, quite remarkably, the neutron-capture cross section could be determined over the full stellar energy range of astrophysical relevance (see Fig.2 in \cite{Marrone04}). This feature could not be matched by any $s$-process branching nuclei measured later due to the much more difficult experimental conditions.

Although the limitations of the phenomenological $s$-process model \cite{Kaeppeler89} were already uncovered many years before after a series of dedicated experiments at the Karlsruhe Van de Graaff accelerator \cite{Arlandini99}, the new n\_TOF results on $^{151}$Sm($n,\gamma$) served to confirm such limitations and helped to refine the physical conditions of modern models of thermally-pulsing asymptotic giant-branch (TP-AGB) stars \cite{Gallino98}. In the context of the classical $s$-process\cite{Kaeppeler89} the large cross section value measured at n\_TOF in combination with the high-neutron density obtained from other neighboring branching nuclei \cite{Best01} would imply an unrealistic temperature regime in excess of 4$\times10^8$ K for the $s$-process operating during He-burning in TP AGB stars. On the other hand, more advanced hydrodynamical models for low-mass TP AGB stars \cite{Gallino98} in combination with the measured cross section helped to constrain the thermal conditions of the He-shell flashes within a range of T$_8$=2.5-2.8 K, as well as to assess more consistently the s- and p-process contributions in the important Sm-Eu-Gd region \cite{Marrone06}.

\subsection*{$^{93}$Zr($n,\gamma$): temperature- and neutron-density conditions in AGB stars}
The measurement of $^{93}$Zr(n,$\gamma$) \cite{Tagliente13} was significantly more difficult despite of its relatively long half-life of 1.6 My (see Fig.\ref{fig:sensitivity}). Indeed, the capture cross-section of this Zr isotope is relatively small ($\sim$96 mb at 30 keV) and the sample enrichment was of only 20\%. Similarly to the measurement of the $^{151}$Sm-branching, two C$_6$D$_6$ detectors were used. In this case they were placed 9 cm upstream in order to reduce the background effect of in-beam $\gamma$-ray Compton scattering in the sample, which has a forward distribution for high-energy $\gamma$-rays. The main limitation in this experiment was due to the background contribution beyond the $\sim$keV neutron-energy range arising from neutrons scattered in the sample and subsequently captured in the surrounding materials (see "setup" labeled spectrum in Fig.1 of Ref.\cite{Tagliente13}). As a consequence of this, the neutron-capture cross section could be determined only up to $\sim$8 keV.

Including the new $^{93}$Zr($n,\gamma$) cross section measured at n\_TOF in state-of-the-art models for thermally pulsing asymptotic-giant branch (TP-AGB) stars \cite{Straniero97} allowed one to get a better insight about the origin of interstellar SiC grains in the CM2 (Mighei type) Murchison chondrite. 
Chondrites are meteorites that have not been modified by melting or differentiation processes in their parent bodies. These chondrites contain SiC grains, individual micro-crystals that were already present in the protosolar nebula and which have survived from destructive processes in the interplanetary disk, in the parent bodies of their host chondrites, and also on their way to the Earth. Utilizing the cross section measured at n\_TOF Lugaro and coworkers \cite{Lugaro03} found that the composition of the SiC grains arises mainly from low-mass (1.5 $M_\odot \leq M \leq 4 M_\odot$) TP-AGB stars, and ruled out the possibility of significant contributions from $\geq 4 M_\odot$ AGB stars, where the $^{22}$Ne($\alpha,n$) source is indeed more efficiently activated than in the lower-mass stars. A series of systematic capture-measurements on the stable $^{90-96}$Zr-isotopes at n\_TOF\cite{Tagliente08a,Tagliente08b,Tagliente10,Tagliente11a,Tagliente11b} in combination with new AGB models \cite{Lugaro04,Karakas09} allowed to explore even more features about the origin of the SiC grains\cite{Lugaro14}. It was found that C-rich 1.25-4 M$_\odot$ AGB stars with metallicities in the range of Z=0.01-0.03 reproduced well the variations of $^{90,91}$Zr/$^{94}$Zr measured in the grains, as well as their Si-isotopic composition. The stellar metallicity was found to be the main parameter for explaining the correlation found between $^{92}$Zr/$^{94}$Zr versus $^{29}$Si/$^{28}$Si in these grains, rather than other effects such as stellar rotation\cite{Lugaro14}. In summary, the complete series of measurements on Zr isotopes allowed one to inspect important open questions about low-mass TP-AGBs, mainly related to the effects of the stellar mass, uncertainties in the $^{13}$C-pocket and the interplay between metallicity and rotation\cite{Lugaro14}. 
Finally, the neutron-energy dependence of the cross-section measured at n\_TOF\cite{Tagliente13}, together with the abundance ratio N(Nb)/N(Zr) observed in extrinsic S-type red-giant stars, allowed one to directly determine an effective value of the $s$-process temperature ($\lesssim$2.5$\times$10$^8$K) in evolved low-mass red-giant stars, independently of stellar evolution models \cite{Neyskens15}. As discussed in the latter work, if new experiments with improved accuracy (or equivalently with a reduced background in the keV neutron-energy region) could be performed, they would help to constrain even more the $s$-process temperature in red-giant TP-AGB stars (see N(Zr)/N(Nb) error bars in Fig.1 of \cite{Neyskens15}).

\subsection*{$^{63}$Ni($n,\gamma$): constraining Cu-Ni-Zn inventory before CCSN explosion}
A similar set-up of two C$_6$D$_6$ detectors placed $\sim$9 cm upstream was used for the measurement of $^{63}$Ni($n,\gamma$) (t$_{1/2}=101$ yr) \cite{Lederer12}. This experiment implied a step further in difficulty (see Fig. \ref{fig:sensitivity}) mostly due to the low sample enrichment ($\sim$10\%) and the reduced number of $^{63}$Ni atoms in the sample. In spite of the excellent TOF (neutron-energy) resolution provided by the 185 m flight-path of EAR1, strong resonance-overlapping contributions in the keV energy range from the dominant $^{62}$Ni content in the sample were difficult to disentangle from the capture-resonances of interest in $^{63}$Ni (see Fig.2 in \cite{Lederer12}). In short, the small cross section ($\sim$67 mb at 30 keV), the reduced number of atoms (10$^{21}$ atoms, 112 mg) and the need to use a PEEK sample encapsulation contributed to a sample-scattered neutron-induced background which, in a similar way as in the measurement of $^{93}$Zr($n,\gamma$), limited the R-matrix analysis of the resolved resonances up to E$_n\sim$10 keV.
The $^{63}$Ni($n,\gamma$) measurement was complemented with the measurement of the stable $^{62}$Ni \cite{Lederer14} for a more reliable evaluation of the impact of these cross sections in the abundance ratio of $^{63}$Cu:$^{65}$Cu in massive stars (M$>$8M$_\odot$), which is now predicted to be 40\% smaller than what was accepted before on the basis of previous cross sections. This result allowed to reduce one of the main uncertainties for the abundances of $^{63}$Cu, $^{64}$Ni and $^{64}$Zn in $s$-process rich ejecta of core collapse supernovae (CCSNe). 

Recently, with the aim of having a fully consistent picture of the Ni-Cu-Zn abundances, a new measurement on  $^{64}$Ni($n,\gamma$)\cite{Tagliente22} has been carried out  at n\_TOF and another one is planned on $^{56}$Fe($n,\gamma$)\cite{Casanovas24}. The measurement of $^{56}$Fe($n,\gamma$) was attempted during the Phase I of n\_TOF \cite{Tain06} and, unfortunately, the in-beam $\gamma$-ray background (discussed in \ref{sec:ntof}) severely constrained the quality of the data. Thanks to the new spallation target design \cite{Esposito21} and the improved experimental conditions at EAR1 this cross section will be measured again in the coming years \cite{Casanovas24}. In the case of $^{64}$Ni($n,\gamma$), the main limitation for a previous measurement was due to the  very low natural abundance (0.92\%) and the limited sample quantity available, which hindered a measurement in the past at EAR1. Thanks to the high flux and improved RF available now at the EAR2 station (20 m flight path) the $^{64}$Ni($n,\gamma$) cross section has been successfully measured in 2024. 

\subsection*{$^{171}$Tm($n,\gamma$): solving rare-earth element anomalies in meteorites}
The difficulty of the $^{63}$Ni($n,\gamma$) measurement was  paralleled afterwards by the neutron-capture measurement of $^{171}$Tm (t$_{1/2}$=1.92 yr) \cite{Guerrero20} (see Fig. \ref{fig:sensitivity}). This experiment was performed at EAR1 utilizing an enlarged setup of four C$_6$D$_6$ detectors for higher detection efficiency. In spite of this, the covered neutron-energy range was limited further down to only $\sim$ 1 keV owing to the much smaller number of atoms in the sample (2$\times$10$^{18}$ atoms). The more limited energy range (1 keV for {$^{171}$Tm versus 10 keV for $^{63}$Ni) reflects that the $^{171}$Tm($n,\gamma$) experiment was indeed even more difficult than the previous one on $^{63}$Ni discussed above. In this case, however, the drawback of the small sample quantity was counterbalanced to some extent by the rather large cross section ($\sim$400 mb at 30 keV), the high sample enrichment (98\%) achieved at PSI via chemical purification and separation and the use of four C$_6$D$_6$ detectors, instead of the two units commonly used before. In addition, this is one of the relatively few cases where the TOF measurement can be conveniently complemented with an activation experiment which, in turn, became crucial for improving the measurement uncertainty down to 10\% at a thermal energy of $kT=30$ keV. The activation experiment was carried out at the SARAF facility \cite{Paul19}. The synergy in this unique combination contributed for building later the new local neutron-activation station (NEAR) at CERN n\_TOF, which is discussed below in Sec.\ref{sec:prospects}. 

The $^{171}$Tm result helped to understand striking isotopic anomalies in rare-earth elements (REE) found in bulk SiC grains of the Murchison meteorite, particularly in three samples where the Yb-isotopic composition was analyzed \cite{Yin06}. The $^{171}$Yb/$^{172}$Yb- and $^{173}$Yb/$^{172}$Yb-isotopic ratios could be well reproduced after the new cross section measured at n\_TOF and detailed modeling of the so-called third dredge-ups (TDUs) included in updated stellar models of low-mass stars \cite{Cristallo11}. SiC grains are thought to be synthesized after TDUs, during which by-products of nuclear burning occurring in stellar interiors, including carbon and s-process isotopes, are transported to the surface of the star. 
Another important consequence of understanding the SiC grain composition is that about one-half of the mass of the carbon-rich envelope is ejected into the interstellar medium (ISM) by winds of low-mass AGB stars has the isotopic composition fixed by the last TDU in such stars \cite{Lugaro03}.

\subsection*{$^{204}$Tl($n,\gamma$): Shedding light on the origin of the heaviest $s$-only nucleus $^{204}$Pb}
$^{204}$Tl (t$_{1/2} =3.78$yr) was the latest and most challenging $s$-process branching measured at EAR1 still with conventional C$_6$D$_6$ detectors \cite{Casanovas24b} (see Fig.\ref{fig:sensitivity}). The sample contained just 2.66$\times$10$^{19}$ atoms of $^{204}$Tl, representing an enrichment of only 4\% with respect to the primary sample isotope $^{203}$Tl. One of the main experimental difficulties was due to a $^{60}$Co-contamination in the sample (373 kBq), which required the unconventional use of a 2 mm thick lead shielding in the front surface of each one of the four C$_6$D$_6$ detectors used for the experiment (see Fig.\ref{fig:ear1}). In addition, a rather high analysis threshold in deposited energy, of 600 keV (more than three times the usual value), was necessary to obtain an optimal signal-to-background ratio. The effect of this cut-off threshold could be nevertheless well accounted for thanks to the methodology developed in Ref.\cite{Abbondanno04}, which enables an accurate treatment of such experimental effects by means of realistic MC-calculations of the capture $\gamma$-ray cascades and detailed modeling of the experimental setup. 

The synthesis of $^{204}$Pb is shielded from $r$-process contributions by its stable isobar $^{204}$Hg and thus, the split of the $s$-process path at $^{204}$Tl is the responsible for the existence of all the $^{204}$Pb that we find in the Solar System today.  $S$-only nuclei are stable isotopes that, like $^{204}$Pb, are entirely produced by the the $s$-process because they are shielded from $r$-process contributions by stable isobars. From $^{70}$Ge to $^{204}$Pb there exist $\sim$30 $s$-only nuclei across the full nuclear chart. $S$-only nuclides are particularly relevant for benchmarking the performance of TP-AGB models at different metallicities, and for assessing the contribution of such stars to the chemical composition of our galaxy \cite{Prantzos20}. Previous studies \cite{Gonzalez13} indicated the possible existence of unknown fragmentation mechanisms in the early Solar System affecting the abundance of the lead isotopes in Ivuna-type Chondrites (CI), with respect to the primordial Solar-System abundances. Supernovae are another hypothetical scenario that could potentially affect the abundance of $^{204}$Pb by means of p-process contributions \cite{Rauscher02,Pignatari16}. As it turns out, addressing the primordial $s$-process origin of $^{204}$Pb is relevant for dating the age of meteorites and their components formed in the first 5 Myr of the Solar System \cite{Connelly08}. In particular, Pb-Pb cosmochronometry represents the state-of-the-art for determining the age of the Solar System by dating Calcium-Aluminum rich Inclusions (CAIs) in primitive meteorites \cite{Connelly12}. Because $^{204}$Pb is the only lead isotope which is exempt of radiogenic contributions from the decay of U- and Th-isotopes its abundance is used for relative normalization of the radiogenic components ($^{206,207}$Pb) in the Pb-Pb clock (see Eqs.(1-3) in \cite{Connelly17}). In this way, $^{204}$Pb  becomes key for the unsurpassed $\sim$0.1-0.2 Myr precision \cite{Amelin09} of this cosmochronometer. New AGB nucleosynthesis calculations based on the cross section measured at n\_TOF delivered $^{204}$Pb abundances fully consistent with the latest solar-system abundance compilation by Lodders \cite{Lodders21}. Within the quoted uncertainties this result ruled out the necessity of invoking fractionation mechanisms or any significant $p$-process contribution to the origin of $^{204}$Pb. Reducing further the uncertainty on the $s$-process contribution to $^{204}$Pb will require an accurate assessment of the thermal dependency of the $\beta$-decay rate of $^{204}$Tl, which could be achieved in the next years in the framework of the PANDORA project \cite{Mascali23}.

Neutron-capture measurements on even more difficult $s$-process branchings utilizing samples with smaller number of atoms, lower enrichment and/or higher $\gamma$-ray activity, required the use and optimization of the second experimental area EAR2 at only 20 m from the spallation source. EAR2 delivers an instantaneous neutron flux, which is more than two orders-of-magnitude larger than the one available at EAR1 \cite{Lerendegui16}, making it a unique tool for suppressing the background contribution arising from the sample radioactivity. In addition, developing two new detection systems was necessary for the measurement of the next two, even more challenging, $s$-process branching nuclides $^{79}$Se and $^{94}$Nb, as described below.

\begin{figure*}[!htbp]
\centering
\includegraphics[width=2\columnwidth]
{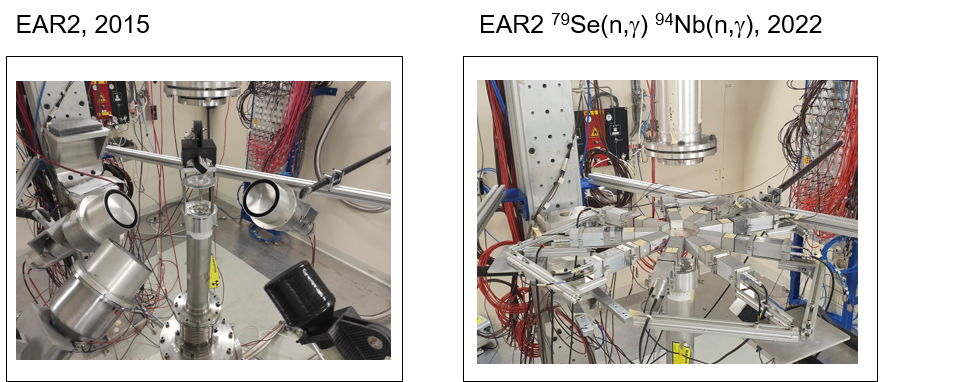}
\caption{Left: Conventional (large-volume) C$_6$D$_6$ detectors for capture measurements in EAR2. Notice that due to their large volume and the high neutron-flux the detectors had to be placed far from the sample in order to limit the count-rate per detector. Right: Upgraded setup with nine small volume C$_6$D$_6$ sTED detectors in a compact configuration surrounding the capture sample.}
\label{fig:ear2}       
\end{figure*}

\subsection*{$^{94}$Nb($n,\gamma$): disentangling Mo-isotopic anomalies in presolar SiC grains}
The sample of $^{94}$Nb (t$_{1/2} = 2\times10^4$ yr) contained 9$\times$10$^{18}$ atoms with only a 1\% enrichment with respect to $^{93}$Nb. In addition, the sample itself contained a $^{60}$Co contamination of 10 MBq. These challenging features, exceeded the experimental capabilities of EAR1 (see Fig.\ref{fig:sensitivity}) and hindered the use of conventional C$_6$D$_6$ detectors (see Fig.\ref{fig:ear2}-left). The capture yield of the $^{94}$Nb($n,\gamma$) reaction could be successfully measured up to 10 keV of neutron energy implementing in EAR2 a new array of nine small-volume C$_6$D$_6$ (sTED) detectors \cite{Alcayne24,Balibrea25} in a compact configuration around the sample (see Fig.\ref{fig:ear2}-right). The high instantaneous flux of EAR2 and the new set-up of detectors optimized for very high count-rate capability became crucial for overcoming the limitations imposed by the $^{60}$Co sample activity\cite{Balibrea24}. 

The neutron capture cross section measurement of $^{94}$Nb at n\_TOF and other stable Mo-isotopes discussed below, was motivated by the better understanding of AGB-con\-tri\-bu\-tions to the composition of mainstream SiC grains. Mainstream grains arise from low-mass C-rich AGB stars, as it is inferred from their isotopic signature \cite{Liu15} and the presence of SiC dust around such stars \cite{Speck09}. The heavy-element isotopic compositions of mainstream grains, like Mo-i\-so\-to\-pes, is less affected by initial stellar composition or GCE effects as compared to light-elements (A$<$56) and, for this reason, these elements are especially well suited for the study of AGB-stellar processess \cite{Liu24}. The Mo-isotopic abundances of mainstream grains were generally well reproduced by state-of-the-art AGB models, with the exception of $^{94}$Mo \cite{Lugaro03}. There are several reasons that could affect the production of $^{94}$Mo in stars, among them the branching at $^{94}$Nb (see Fig.1 in \cite{Lugaro03}) leading to the formation of $^{94}$Mo. This contribution is enhanced during the He-shell flashes of TP-AGB stars by the reduced $^{94}$Nb half-life of only a few days at T=3$\times$10$^8$ K \cite{Takahashi87}. Uncertainties on the thermal-dependency of the $^{94}$Nb $\beta$-decay rate, and the unknown $^{94}$Nb($n,\gamma$) neutron-capture cross section could be responsible of the observed discrepancy \cite{Lugaro03}. Low-mass TP-AGB models \cite{Lugaro03} predict about 4 times less $^{94}$Mo than the relative quantities found in SiC grains (see Fig.8 in \cite{Lugaro03}). More recent low-mass AGB models with updated magneto\-hydro\-dynamics-based mixing schemes \cite{Vescovi20,Busso21} also yield an anomalous difference between predicted and measured $^{94}$Mo abundance ratios in mainstream-SiC grains (see Fig.8 in \cite{Palmerini21}). Among the different isotopes involved, obviously neutron-capture on the unstable $^{94}$Nb is the most difficult one to measure owing to the reasons highlighted above. However, uncertainties in the cross sections of the stable Mo-isotopes should not be ruled  out and, for that reason, additional measurements on the stable $^{94-96}$Mo isotopes have been recently carried out at CERN n\_TOF EAR1 \cite{Mucciola22,Mucciola23}. Both the data-analysis of $^{94}$Nb($n,\gamma$) and $^{94-96}$Mo($n,\gamma$) are pre\-sen\-tly in progress and, once finalized, the uncertainties from the neutron-capture input data will be removed from this problem. It remains to be investigated if a different thermal dependency of the $^{94}$Nb $\beta$-decay rate in the stellar plasma conditions could have a significant impact. Fortunately, the measurement of this decay-rate at high (stellar) temperature is precisely one of the first priorities of the future PANDORA project \cite{Mascali23}.

\subsection*{$^{79}$Se($n,\gamma$): temperature in Massive and AGB stars}

Following the defined FOM (\ref{eq:fom}), the most challenging $s$-process branching measured so far at CERN n\_TOF via the time-of-flight technique is $^{79}$Se (see Fig.\ref{fig:sensitivity}). $^{79}$Se has a terrestrial half-life t$_{1/2}$ = 3.27(8)$\times$10$^5$ years \cite{Bienvenu07} and the chemical element selenium itself has a melting point of only 494 K. Because of such low melting point an eutectic lead-selenide (PbSe) alloy \cite{Chi22}, highly enriched in $^{208}$Pb and $^{78}$Se, was prepared at PSI with a total mass of 3.9 g\cite{Chiera22} for subsequent activation at the high-flux reactor of ILL-Grenoble. The eutectic alloy was necessary to comply with safety regulations at ILL. The resulting activated sample contained only 2.7 mg of $^{79}$Se, thus corresponding to an enrichment factor as low as 7$\times$10$^{-4}$. In addition, the activity due to the $\gamma$-ray emitting decays of $^{75}$Se and $^{60}$Co from ineluctable impurities activated in the sample were of 5 MBq and 1.4 MBq, respectively.
The complex data-analysis related to the use of the i-TED and sTED detectors in two different experimental areas is expected to be completed soon. We clearly observe a number resonances below $\sim$1 keV related to neutron-capture on $^{79}$Se, which means that it will be possible to provide the very first neutron-capture nuclear-input required for a proper interpretation of this important $s$-process branching point.

\begin{figure}
    \resizebox{0.5\textwidth}{!}{\includegraphics{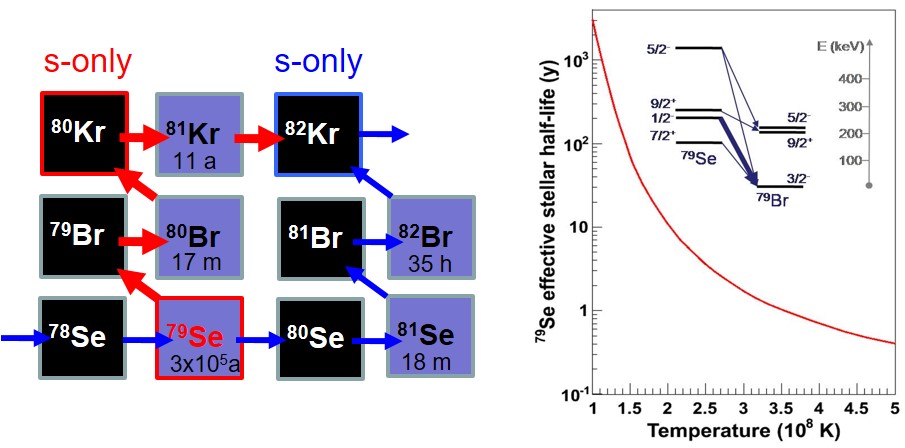}
}
\caption{Left: Part of the nuclear chart showing the $s$-process path at the branching in $^{79}$Se. The $\beta$-decay towards $s$-only $^{80}$Kr is strongly enhanced at higher stellar temperatures (red arrow)  due to the population of the $1/2^-$-isomeric state at low energy. Right: Effective half-life of $^{79}$Se as a function of stellar temperature. The inset indicates levels involved.}
\label{fig:se}       
\end{figure}

The branching at $^{79}$Se represents indeed one of the best cases to constrain the thermal conditions of the $s$-process \cite{Kaeppeler11,Walter86}. Under stellar conditions, the half-life of $^{79}$Se decreases to less than a year due to the population of the isomeric state at only 95.7 keV \cite{Klay88,Takahashi87} (see Fig.\ref{fig:se}). Remarkably, this subtle nuclear-structure effect leads to final abundances for the $s$-only $^{80,82}$Kr isotopes, which are strongly dependent upon the thermal conditions inside the star and the neutron-capture rate at $^{79}$Se. Moreover, this neutron-capture cross section measurement could provide valuable information about $s$-process nucleosynthesis both in massive stars and in AGB stars because the $s$-process path induced by $^{79}$Se is located in the transition region between weak (massive stars) and the main (low-mass AGBs) $s$-process contributions \cite{Walter86}. The stable end-products of the Se branching, the (s-only) Kr isotopes, have been well characterized in presolar graphite grains of the CM2 Murchison chondrite \cite{Amari14}. A striking property of these graphite grains is that their isotopic features depend on their density and, in fact, low-density grains are inferred to predominantly arise from CCSN explosions of massive stars \cite{Amari95a,Amari95b}, while high density ($>$2.1 g/cm$^3$) graphite grains are thought to mainly originate from C-rich AGB stars\cite{Amari14}. Therefore, once the experimental result from the neutron-capture cross section measurement carried out at n\_TOF becomes available in the near future, a complete assessment of the thermal conditions both in AGB and massive stars will become possible by revisiting the readily available Kr-isotopic abundances from presolar low- and high-density graphite grains extracted from the Murchison meteorite.

\section{Recent measurements on stable isotopes}\label{sec:stable}
Although neutron-capture cross-section measurements virtually exist for all stable isotopes of $s$-process interest, the related uncertainties are in most cases far from the $\pm$5\% uncertainty level required for $s$-process model calculation utilizing low-mass AGB models \cite{Cristallo11,Cescutti18,Cescutti19} and massive-star models \cite{Nishimura18a,Nishimura18b}. Certainly, a large effort has been made over the last years in order to access the challenging unstable $s$-process nuclei. 
However, it is worth emphasizing that many important facets of stellar nucleosynthesis in AGB- and massive stars still require significant effort with stable isotopes to reduce present uncertainties. This effort should focus on new measurements extending across the full energy range of astrophysical interest. In the following some recent examples are discussed in more detail.

\subsection{$S$-process bottlenecks: $^{140}$Ce($n,\gamma$) and $^{209}$Bi($n,\gamma$)}\label{sec:bottlenecks}
The neutron shell-closure effects along the $s$-process path are reflected in small capture cross sections, which lead to characteristic abundance peaks in the mass-regions of Sr, Y, Zr (N=50), Ba, La, Ce, Nd, Sm (N=82) and Pb, Bi (N=126). Because $s$-process bottleneck elements, in general, show-up prominently in spectroscopic observations of stellar atmospheres, from the observational point of view they can be accurately quantified in different types of stars, thus becoming good candidates for testing the details of the stellar models.
However, cross section uncertainties for a large number of neutron-magic nuclei are still very sizable (see Fig.6 in \cite{Dillmann23} and Fig.1 in \cite{Domingo23b}). 
An uncertainty below $\sim$2\% \cite{Kaeppeler11,Kaeppeler11b} would be more convenient for a proper interpretation of the observed abundances and for exploring different aspects of low-mass AGB stars, such as the effect of metallicity \cite{Bisterzo10a} or the amount of $^{13}$C in the pocket \cite{Bisterzo10b,Gallino98,Arlandini99}.
In practice, experiments with neutron-magic nuclei often present significant challenges due to their extremely low radiative capture cross sections. In some cases, these cross sections barely exceed a few millibarns at $kT = 30$ keV\cite{Domingo06a}. Consequently, the radiative capture cross section can be up to three to four orders of magnitude smaller than the competing elastic scattering one within the stellar energy range of interest.
In these scenarios, the high number of scattered neutrons may interact (promptly or after partial thermalisation) with materials in the surroundings of the experimental setup and detectors themselves. Such contaminant interactions emit radiation that contaminates the measurement and increases background levels. This phenomenon, referred to as neutron sensitivity, has driven advancements in detection systems. C$_6$D$_6$ detectors were introduced as a more suitable alternative to the C$_6$F$_6$ detectors commonly used in earlier studies \cite{Corvi88}. At n\_TOF, carbon-fibre based C$_6$D$_6$ scintillation detectors were initially developed \cite{Plag03} and progressively improved after many years of operation\cite{Mastinu21}, thereby aiming at minimizing potential neutron sensitivity backgrounds (see Fig.\ref{fig:ear1}). These developments were of particular importance for achieving a high systematic accuracy in the measurement of neutron-magic isotopes \cite{Domingo06a}. 
Another source of background that may significantly affect these measurements is the in-beam $\gamma$-rays background, which becomes enhanced when measuring high-Z samples. In summary, the very-low capture cross sections of bottleneck isotopes makes their measurement very sensitive to any background contribution, and especial care needs to be taken in the design of the detection system and in the detectors themselves. Two recent $s$-process bottelneck measurements will be discussed here: $^{140}$Ce\cite{Amaducci24} and $^{209}$Bi\cite{Balibrea23b}.

\subsection*{$^{140}$Ce: the second $s$-process peak}
For the measurement of the $^{140}$Ce($n,\gamma$) cross section \cite{Amaducci24} a highly enriched $^{140}$Ce sample was produced at PSI. The sample mass was of 12.318 g (Ce-oxide) and the amount of $^{142}$Ce was of only 0.6\%. 
This enabled a high-resolution measurement at EAR1 covering a large neutron-energy range (up to 65 keV), thereby utilizing the conventional setup of four C$_6$D$_6$ detectors optimized for low neutron sensitivity backgrounds. The resulting Maxwellian averaged cross section was up to 40\% higher than previously reported values and, in combination with low-mass AGB stellar models \cite{Straniero06,Cristallo11} a reduction of 20\% in the $s$-process abundance of $^{140}$Ce was obtained with respect to previous estimations.

The $^{140}$Ce($n,\gamma$) experiment was motivated by a discrepancy of 30\% found between the Ce abundance predicted with AGB models and the quantity determined from spectroscopic observations of stars in the M22 globular cluster, for which the $s$-process pollution could be well isolated \cite{Straniero14}. Interestingly, a good agreement was found between abundance predictions and observations for the neighbouring bottleneck elements: Ba, La, Nd and Sm (see Fig.11 in \cite{Straniero14}). The new cross section leads to an even smaller predicted Ce-abundance (see Fig.1 in \cite{Amaducci24}), which does not solve this discrepancy and makes this problem even more intriguing. In addition, other independent activation experiments \cite{Sahoo24} provide a MACS value which differs significantly from the n\_TOF result. This situation reflects the inherent difficulty in the measurement of bottleneck isotopes, which are characterized by their very low cross sections. As a consequence, a new activation experiment has been carried out at the new NEAR facility (see Sec.\ref{sec:prospects}) of n\_TOF utilizing the same high-quality sample that was employed for the TOF experiment. 
For completeness, it is worth mentioning that some impact on the observed abundance may come from possible $i$-process contributions \cite{Cowan77,Denissenkov21}, that could be enhanced by a rather small $^{140}$Ba($n,\gamma$)  cross section. Obviously, this contribution cannot resolve the discrepancies in the MACS determined from different experiments. Both theoretical, experimental and (probably) observational efforts seem to be required at this stage in order to shed some light on this interesting case. 

\subsection*{$^{209}$Bi: the third $s$-process peak}
Bismuth is the heaviest element produced by the $s$-process in low-mass AGB stars, it is monoisotopic (A=209), and its abundance in the Solar System is relateively well known \cite{Lodders21}.
The $^{209}$Bi($n,\gamma$) cross section was measured at CERN n\_TOF EAR1 in 2001\cite{Domingo06a} with a set-up similar to the one shown in Fig.\ref{fig:ear1}-left. Together with the measurement of the neighboring lead isotopes $^{204,206,207}$Pb\cite{Domingo06b,Domingo07a,Domingo07b}, relevant information could be obtained for the study of the termination region of the $s$-process path (see previous references and Sec.5 in Ref.\cite{Massimi22}). In particular, rather accurate abundance constraints could be derived for the $r$-process contribution to $^{209}$Bi (see Fig.5 in \cite{Massimi22}) via the so-called $r$-process residuals ($N_r = N_\odot - N_s$), given that the corresponding $s$-process abundance was well determined from the measured cross section and TP-AGB stellar models (see \cite{Domingo06a} and references therein) and the fact that the solar system abundance of Bi ($N^{Bi}_\odot$) is relatively well known ($\sim$7\% relative uncertainty) from the isotopic analysis of CI-chondrites \cite{Lodders21}. In turn, the $r$-process residuals analysis helped to qualify or reject some of the former nuclear-mass models utilized in $r$-process model calculations \cite{Domingo06c,Massimi22}.

\begin{figure}
    \resizebox{0.4\textwidth}{!}{\includegraphics{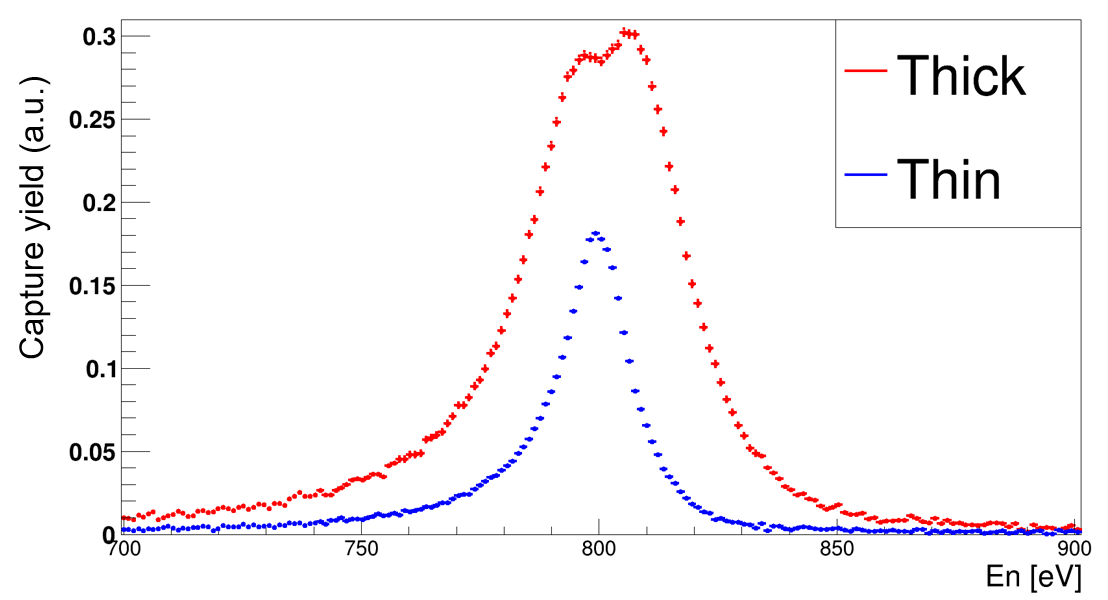}
}
\caption{Comparison of the capture yields measured at n\_TOF EAR2\cite{Balibrea23b} for the first capture-resonance in $^{209}$Bi+n with a thick and a thin sample.}
\label{fig:bi}       
\end{figure}

However, despite efforts in reducing the intrinsic neutron-sensitivity of the detection apparatus, the first measurement of $^{209}$Bi at n\_TOF encountered many difficulties due to the aforediscussed in-beam $\gamma$-ray background, which was significantly enhanced by the high-Z (83) of the sample. This background was indeed largest in the $\sim$10 keV neutron-energy range, which hindered the observation of some resonances and reduced significantly the statistical accuracy in many of them.
With the new improved experimental conditions discussed in Sec.\ref{sec:ntof} a new measurement of $^{209}$Bi($n,\gamma$)  was successfully carried out in 2024 at EAR2\cite{Balibrea23b}. Preliminary results indicate that with the new measurement one could obtain an improvement in both the covered energy range and the number (and statistical accuracy) of resonances observed. In addition, the high-flux at EAR2 enabled the measurement of two independent Bismuth samples, with two different thicknesses, which permitted a better assessment of multiple-scattering and neutron-sensitivity effects. These two experimental effects are of particular importance in the case of broad s-wave resonances (see Fig.\ref{fig:bi}).

\subsection{$^{28-30}$Si($n,\gamma$): disentangling Si-Ti-Mg isotopic anomalies in mainstream SiC grains}\label{sec:si}
Even for the most abundant and best known mainstream SiC presolar grains, already discussed here in the context of the $^{94}$Nb-branching in Sec.\ref{sec:branchings}, the mass and metallicity properties of their progenitor AGB stars are still quite ambiguous \cite{Cristallo20,Lugaro18}. Modern galactic chemo-dynamical models indicate that $M \sim 2 M_\odot$ and $Z\sim Z_\odot$ C-rich AGB stars seem to be the main progenitors of these SiC grains in our Solar System \cite{Cristallo20}, whereas previous studies rather point to AGB stars with masses up to 4 $M_\odot$ and $Z\sim 2 Z_\odot$ \cite{Lugaro18}. In turn, the mass- and metallicity-distributions of the AGB-star population is of fundamental relevance for $s$-process nucleosynthesis and for understanding the chemical evolution of our galaxy. As discussed before in the context of the $^{94}$Nb branching in Sec.\ref{sec:branchings}, heavy element isotopic compositions of mainstream grains, like Zr or Mo, can help to avoid effects related to the ISM composition or the imprints of GCE. In practice, improving current uncertainties in the neutron-capture cross sections of the light Si-isotopes, commonly used for the isotopic-classification and interpretation of SiC grains\cite{Liu24}, may also contribute to disentangle possible imprints of the ISM composition in the formation of these grains from the intrinsic nucleosynthesis contribution of the C-rich AGB progenitor.

Experimentally, neutron-capture measurements on silicon isotopes are challenging owing to their very small neutron capture cross section. A further difficulty is the significant direct capture contribution expected in these light nuclei, which can only be estimated theoretically or inferred from the thermal neutron-capture value if it is precisely known. Latest neutron-capture measurements on the Si-isotopes were rather inconclusive regarding the interpretation of the isotopic-composition of mainstream SiC grains, with important shifts with respect to expectations found for $^{29,30}$Si \cite{Guber03}.  
In an effort to improve this situation a new series of measurements on $^{28,29,30}$Si$(n,\gamma)$ were carried out at n\_TOF utilizing highly enriched samples, state-of-the-art C$_6$D$_6$ detectors and both EAR1 and EAR2 experimental areas\cite{Lederer23}.  The two independent measurements in EAR1 and EAR2 with complementary setups (four C-fiber C$_6$D$_6$ and nine sTED-detector array) allows one to keep under control systematic effects related to multiple-scattering and other experimental effects. For $^{30}$Si($n,\gamma$)  resonances have been observed up to several hundreds keV and the MACS at 30 keV seems to be dominated by the first strong resonances in the 5-15 keV energy range. Fig.\ref{fig:si} displays an example of the excellent statistical quality of the data, showing a capture yield that is remarkably smaller than the values predicted by evaluations.
\begin{figure}
    \resizebox{0.5\textwidth}{!}{\includegraphics{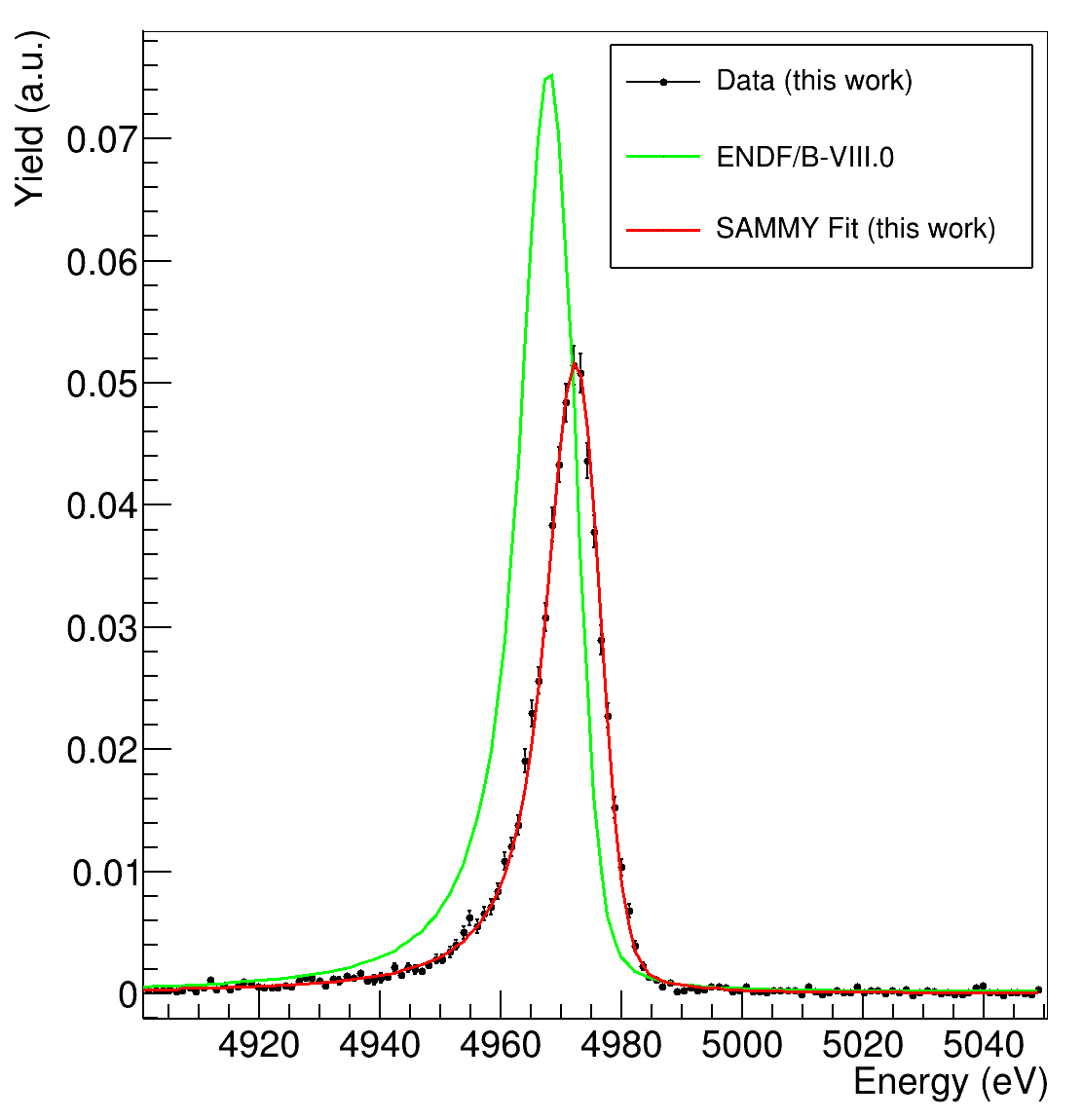}
}
\caption{R-matrix fit and its expected shape based on the ENDF/B-VIII.0 parameters of the first resonance in $^{30}$Si+n measured in EAR1.}
\label{fig:si}       
\end{figure}

\section{Present limits and new horizons}\label{sec:prospects}

Some of the main persistent issues and limitations in the field of neutron-capture experiments for $s$-process studies are discussed in this section. A synopsis is also given on current plans for overcoming such limitations in the coming years at n\_TOF, as well as new envisioned ideas for advancing further in more longer-term plan (coming decades).

The very first limitation in most of the challenging cases shown in Fig.\ref{fig:sensitivity} starts with the quality of the capture sample itself. In spite of efforts and resources invested in manufacturing the best suitable samples for capture experiments, in most cases their final quality and composition still represents one of the main restrictions. Let us take as example the most challenging $^{79}$Se($n,\gamma$) case discussed in Sec.\ref{sec:branchings}. Having a chemically (Se) pure sample, instead of a lead-selenide eutectic alloy, would represent a remarkable advantage. All backgrounds arising from neutron-scattering in the lead content would be removed and also, the $^{60}$Co contamination in the measured sample (1.4 MBq) arose from traces of $^{nat}$Co in the highly enriched $^{208}$Pb raw material used for producing the eutectic PbSe alloy \cite{Chiera22}. It is technically feasible to obtain a radiochemically pure sample of Selenium, such as the sample prepared at PSI \cite{Jorg10} for the measurement of the $^{79}$Se half-life \cite{Bienvenu07}. However, the cost and resources required to obtain a sufficient number of $^{79}$Se atoms ($>$10$^{18}$) for a TOF experiment would be prohibitive.
A radio-isotopically pure sample of $^{79}$Se would enable a further huge improvement in reducing backgrounds during the measurement, and probably would lead to an experimental determination of the neutron-capture cross section in the full energy range of astrophysical interest ($\sim$1 eV -- 100 keV).  This additional purification step would require the use of a dedicated mass-separator and the final sample amount would strongly depend on the design of such apparatus and the initial quantity and quality of material available.
Only a few laboratories worldwide are equipped with the instrumentation and knowledge required to handle and produce such samples \cite{Schumann16}. A big coordination effort is being pursued in the context of the SANDA \cite{sanda} 
and its follow-up APRENDE project \cite{aprende}, as well as by the International Nuclear Target Development Society \cite{Schumann23}.
An off-line isotope separator for such purpose has been envisaged at PSI, which would be well suited for the production of radio-isotopically pure samples with sufficient amounts, in suitable backings and free from oxides (see also related discussion in Ref.\cite{lrp24}). 

One of the main limitations in TOF experiments that becomes apparent after analyzing the seven $s$-process branching examples discussed in Sec.\ref{sec:branchings} (see Fig.\ref{fig:sensitivity}) is the neutron-energy range efficiently covered in such experiments. The white neutron-flux spectrum and low duty-cycle of n\_TOF allows one to fully cover the 1 eV to 100 keV energy range of interest for $s$-process studies. However, that wide energy range could only be analyzed in the very first $^{151}$Sm($n,\gamma$)  experiment (see Fig.1 in Ref.\cite{Marrone04}). For all other $s$-process branching cases discussed in Sec.\ref{sec:branchings}, the upper data-analysis limit was between 1 keV and 10 keV, thus missing an important (if not the most important) part of the stellar spectrum. In many cases the energy range beyond 10 keV corresponds already to the unresolved resonance region (URR) and thus, if the signal-to-background ratio is not adequate, it becomes increasingly difficult and uncertain to reliably extract the cross section. In such cases average resonance parameters were determined in the measured (low-energy) range, and a simulation of random-resonance sequences was computed to determine the MACS at higher $kT$ thermal energies (see e.g.\cite{Guerrero20,Casanovas24b}).

Improving the experimental background conditions, or enhancing further the sensitivity of the detection apparatus with ideas similar to i-TED \cite{Domingo16,Babiano21} or sTED \cite{Alcayne24} could help to overcome this limitation in future experiments. Presently, there are several efforts in this direction. A new (enlarged) version of the sTED array is being developed for experiments at both EAR1 and EAR2, thereby exploring a new technique with intermediate detection efficiencies \cite{Mendoza23}. Moreover, a new version of the sTED-detector array based on stilbene-d$_{12}$ organic crystals instead of C$_6$D$_6$ cells \cite{Alcayne24}, the so-called Stilbene-d$_{12}$ deTector ARray (STAR), has been designed and is already under construction \cite{Balibrea25}. The use of even smaller detection volumes ($25\times25\times50$ mm$^3$) of higher intrinsic detection efficiency than liquid C$_6$D$_6$ is expected to enhance further the signal-to-background ratio in future experiments at EAR2. It is foreseen that STAR will be commissioned in 2026 and operational for first capture experiments in 2028, after the third long shutdown (LS3) of CERN.
\begin{figure*}[!htbp]
\centering
\includegraphics[width=2\columnwidth]
{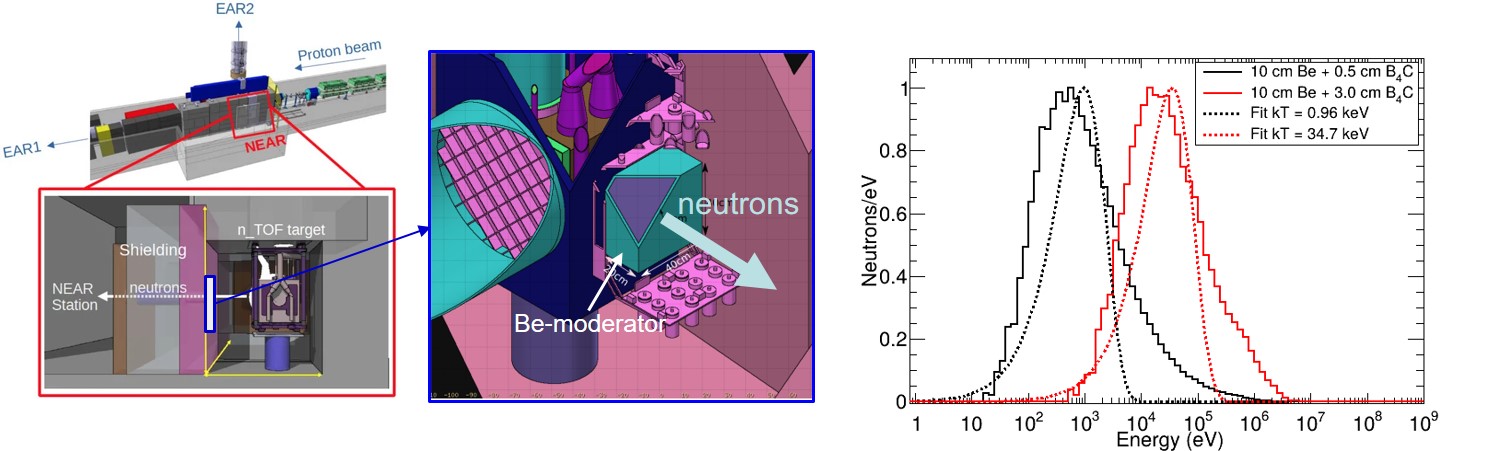}
\caption{Drawings showing the location and different elements of the new NEAR station (left) and some quasi-Maxwellian neutron distributions obtained via MC-Simulation for different B$_4$C- filter thicknesses, leading to mean kT values of about 1 keV and 35 keV.}
\label{fig:near}       
\end{figure*}
A complementary and alternative effort, that in some specific cases may help to overcome the aforementioned limitation in neutron-energy range is the new NEAR activation station \cite{Stamati23,Gervino22}, located at only 3 m from the spallation source (see Fig.\ref{fig:near}). With an average flux of $\sim$5$\times$10$^7$ n/cm$^2$/s this new facility will open the possibility to conduct, when feasible, activation experiments with quasi-Maxwellian neutron distributions spanning from $\sim$keV up to $\sim$100 keV. The precise value of $kT$ will be determined by the chosen (variable) thickness of a boron carbide (B$_4$C) filter embedding the sample under study \cite{Stamati23}. A commissioning of the NEAR station and a flux characterization has been already carried out \cite{Stamati23} and its main features and ancillary equipment are described in detail in Ref.\cite{Patronis22}. A Be moderator directly attached to one side of the spallation target will be installed during 2025, which will allow to improve the quality of the quasi-Maxwellian neutron spectrum significantly (see insets in Fig.\ref{fig:near}). 

The potential of NEAR for new activation measurements on unstable nuclei will be boosted by the availability of the nearby ISOLDE facility and the large yields that can be produced therein \cite{Zanini14,Turrion08}. Radioisotopically pure samples with sufficient number of atoms ($\sim$10$^{15-16}$) will be produced at ISOLDE for the envisaged activation experiments at n\_TOF NEAR. At present, this approach is already being tested with the production of a $^{135}$Cs sample at ISOLDE utilizing General Purpose Separator (GPS) \cite{Lerendegui22b}. Once the sample becomes available, the activation measurement at NEAR is foreseen for 2026 \cite{Lerendegui24}. 
For the measurement of isotopes leading to short-lived activated products the CYCLING (CYCLIc activation station for (N,G)  experiments) is being prepared and background characterization measurements at NEAR have been already carried out utilizing active (NaI-scintillation) detectors \cite{Lerendegui22a}.
With NEAR and NEAR-CYCLING, in combination with the enhanced sample-production capabilities of ISOLDE, ILL and PSI, we expect to expand the number of ($n,\gamma$)  cases of astrophysical interest in the coming years, as well as to provide complementary information in the 10-100 keV energy range, for those TOF experiments in EAR1 or EAR2 that are technically restricted to maximum neutron-energies of $\sim$10 keV. Several $s$-process cases to be tackled in the future at NEAR have been discussed already in Ref.\cite{Lerendegui23a,Domingo23b}, including some direct measurements for $i$-process studies.

In the longer term, the n\_TOF collaboration is planning to expand further on the large potential of the activation technique and on its complementarity for many TOF experiments in EAR1 and EAR2. In this respect, an expression of interest for a new high-flux activation station n\_ACT has been prepared \cite{Lederer24}. The latter would be built at the SPS Beam Dump Facility (BDF) \cite{BDF22}, where  ultra-high neutron fluxes will be produced. n\_ACT would allow to strategically profit such high neutron fluence. Preliminary designs for n\_ACT \cite{Lederer24} predict about three orders of magnitude higher fluxes than what is presently achievable at NEAR.
Again, in combination with radio-isotopically pure samples from ISOLDE, this new facility may open-up the possibility to access for the first time direct (activation) measurements on samples available in minor quantities, including some neutron-rich unstable nuclei in the $i$-process path.

Finally, regarding possible long-term advancements, it is worth discussing here the concept of direct neutron-capture  measurements in inverse kinematics with a stationary neutron target, which was proposed in Refs.\cite{Reifarth14,Reifarth17}. If feasible, this novel methodology will allow one to overcome experimental difficulties that have been discussed above, mainly those connected with the production of the capture sample, the relatively large number of atoms required, the half-life of the isotope of interest and the radioactivity of the sample itself. Other additional aspects related to detector-sensitive backgrounds could be significantly improved as well. This new technique involves the combination of three main elements, namely, a radioactive ion-beam facility (preferably ISOL-type for low-energy secondary beams in the stellar energy range of interest 50-500 keV), a suitable low-energy storage ion-ring, and a heavily moderated spallation neutron source to drive the free neutron-gas target. A neutron target demonstrator is already being developed at Los Alamos Neutron Science Center (LANSCE) \cite{Cooper24} and there are plans to demonstrate its performance in the coming years with a proof-of-concept experiment utilizing single-pass stable-beam experiments \cite{Cooper24}. At TRIUMF-Vancouver, there are also plans to exploit the high-intensity ISOL yields of the ARIEL facility in combination with a low-energy storage ring (TRISR) and an alternative option based on compact neutron generators for the static neutron target is under study \cite{Dillmann23}. 

Interestingly, most of the three basic elements required for the new methodology are readily available at CERN, or at least a large expertise exists with them. On the one hand, ISOLDE would be an ideally well suited installation for the production of low-energy neutron-rich nuclei with high yields for this application. On the other hand, the 25 years of n\_TOF experiment have led to a great expertise in the design and operation of spallation-targets \cite{Esposito23} (see Fig.\ref{fig:target}), including also the new plans at the SPS-BDF facility \cite{Lederer24}. A low-energy storage ring, the Test Storage Ring (TSR) was built \cite{Grieser12} and operated at MPI-Heidelberg, with later (unsuccessful) plans to install it at HIE-ISOLDE \cite{Butler16}. More recently, in the framework of the EPIC ISOLDE upgrade, there is a renewed interest for a new Isolde Storage Ring (ISR), which could become a reality at CERN in the coming decade. 

In summary, after 25 years of stunning developments and fascinating results at CERN n\_TOF, new disruptive approaches like the inverse-kinematics concept could represent in the long-term future a new paradigm in this field. This new methodology could open the possibility of measuring the few remaining $s$-process branching nuclei, most of the isotopes in the $i$-process path and even approach the neighborhood of the $r$-process path. 

\section*{Acknowledgments}
CDP acknowledges Nan Liu (Boston University) for very helpful comments and suggestions.
The authors acknowledge support from all the funding agencies of participating institutions. Part of this work was supported by the European Research Council (ERC) under the European Union's Horizon 2020 research and innovation programme (ERC-COG Nr.~681740 and ERC-STG Nr.677497), European H2020-847552 (SANDA), the MCIN/AEI 10.13039/\-501100011033 under grants Severo Ochoa CEX\-2023-001\-292\--S, PID2022-138297NB-C21, PID2019-104714GB-C21, FPA\-2017-83946-C2-1-P, FIS2015-71688-ERC, FPA\-2016-77689-C2-1-R, RTI2018\--098117\--B\--C21, PGC2018\--096717\--B\--C21, PID2021\--123100NB\--I00 funded by MCIN/AEI 10.13039/\-501100011033/\-FEDER, UE, PCI2022\--135037\--2 funded by MCIN/\-AEI 10.13039/\-501100011033 and EU-\-NextGenerationEU/\-PRTR.

For the purpose of open access, the corresponding author has applied a Creative Commons Attribution (CC BY-NC-ND 4.0) license to any Author Accepted Manuscript version arising from this submission.

%
%
\bibliography{bibliography}

\end{document}